\documentstyle[epsf,epsfig,psfig,12pt]{article}
\newcommand{\BE}{\begin{eqnarray}}
\newcommand{\EE}{\end{eqnarray}}
\newcommand{\BF}{\begin{figure}[htbp]}
\newcommand{\EF}{\end{figure}}
\newcommand{\BC}{\begin{center}}
\newcommand{\EC}{\end{center}}
\newcommand{\order}{\mbox{${\cal O}$}}
\catcode`\@=11
\def\makepreprititle{\par
  \begingroup
  \def\thefootnote{\fnsymbol{footnote}}
  \def\
@makefnmark{\hbox 
  to 0pt{$^{\@thefnmark}$\hss}} 
  \if@twocolumn 
  \twocolumn[\@makepreprititle] 
  \else \newpage
  \global\@topnum\z@ 
  \@makepreprititle \fi\thispagestyle{empty}\@thanks
  \endgroup
  \setcounter{footnote}{0}
  \let\makepreprititle\relax
  \let\@makepreprititle\relax
  \gdef\@thanks{}\gdef\@author{}\gdef\@title{}
  \gdef\@preprintnumber{}\gdef\@preprintdate{}\gdef\subtitle{}
  \let\thanks\relax}
\def\preprintnumber#1{\gdef\@preprintnumber{#1}}
\def\preprintdate#1{\gdef\@preprintdate{#1}}
\def\subtitle#1{\gdef\@subtitle{#1}}
\def\@makepreprititle{\newpage
{\def\baselinestretch{1}
  \begin{flushright} \@preprintnumber \par
  \@preprintdate \end{flushright} } \par
  \begin{center}
\vskip 1.5em
  {\LARGE \@title \par} \vskip 2.5em 
  {\large \lineskip .5em
  \begin{tabular}[t]{c}\@author 
  \end{tabular}\par}
  \vskip 1em {\large \@date} \end{center}
  \par
  \vfil} 
\date{\sl Department of Physics, Tohoku University\\Sendai, 980 Japan}
\preprintdate{~}
\preprintnumber{~}
\subtitle{}
\def\abstract{\if@twocolumn
\section*{Abstract}
\else \normalsize 
\begin{center}
{\bf Abstract\vspace{-.5em}\vspace{0pt}} 
\end{center}
\quotation 
\addtocounter{page}{-1}
\fi}
\def\endabstract{\if@twocolumn\else\endquotation\fi}
\def\spacing#1{\def\baselinestretch{#1}
\typeout{baselinestretch is modified to \baselinestretch}}
\catcode`\@=12

\spacing{1.3}
\newcommand{\lsim}{\mbox{ \raisebox{-1.0ex}{$\stackrel{\textstyle <}
{\textstyle \sim}$ }}}
\newcommand{\GeV}{\mbox{GeV}}
\newcommand{\TeV}{\mbox{TeV}}
\renewcommand{\thefootnote}{\fnsymbol{footnote}}
\makeatletter
\long\def\@makefntext#1{\parindent 1em\noindent 
 \hbox to 1.8em{\hss${\@thefnmark}$ }#1}
\def\@makefnmark{\hbox{$^{{\@thefnmark}}$}}
\makeatother
\preprintnumber{TU-503}
\preprintdate{May 1, 1996}
\title{Top Mass and Isospin Breaking in\\
Dynamical Symmetry Breaking Scenario}
\author{T.~Asaka, 
Y.~Shobuda, 
and Y.~Sumino\\
\\
\sl Department of Physics, Tohoku University\\
\sl Sendai, 980-77 Japan}
\date{~}
\begin{document}
\makepreprititle
%
%
\begin{abstract}

We consider a scenario where the top-quark mass is generated
dynamically, and study the implication of the present experimental
values for $m_t$ and the $T$ parameter.
We assume technicolor-like scenario for inducing the $W$ mass and 
an effective four fermi operator for inducing the top-quark mass.
We also
assume that only this four fermi operator is relevant at low energy.
Then we estimate in detail the strength $G$ and the
intrinsic mass scale $M$ of the four fermi operator.
Unitarity bound is used 
to quantify the strength of $G$.
We find that 
$G/4\pi \sim {\cal O}(1)$ and that $M$ is of the order of
$\Lambda_{TC} \simeq 1 \sim 2$~TeV or less.
Namely the four fermi operator cannot be treated as
`point-like' around the electroweak scale.
Furthermore we estimate the contribution of the four fermi operator to
the $T$ parameter.
We find that the QCD correction to the top-quark mass function
reduces the contribution to the $T$ parameter by about 40\%.
By comparing the results 
with the present experimental bound, we obtain another upper 
bound on $M$ which is typically in several to 10~TeV region.

\end{abstract}
\clearpage
\section{Introduction}
%
%

The $SU(2) \times U(1)$ gauge theory for describing the electroweak
interactions has been very successful both theoretically and
experimentally.  However, all the experimental tests have been done 
for its gauge part and  we have little knowledge on 
the electroweak symmetry breaking mechanism so far.
In the light of the naturalness problem,
we may suppose that there exists some new physics 
related to the electroweak symmetry breaking at the energy scale
100~GeV $\sim$ 1~TeV.
Dynamical symmetry breaking is one of the attractive candidates
for the solution to the naturalness problem.
We consider this possibility and study the implication of the present
experimental data for the top-quark mass\cite{mt} 
and the $T$-parameter\cite{STU,t,Matsumoto}.

In dynamical symmetry breaking scenarios such as technicolor models, an 
effective four fermi operator 
\BE
{\cal L} ~ = ~ \frac{1}{\Lambda^2} \overline{U_L} U_R 
\overline{t_R} t_L + h.c.
\label{in4fint}
\EE
is introduced
in order to generate the top-quark mass,
where $\Lambda$ represents the new physics scale (extended technicolor
scale) 
and $U$ denotes a new fermion (techni-fermion)
introduced in the symmetry breaking sector.
When this fermion forms a pair condensate
$\left<\overline{U_L}U_R\right> \neq 0$,
the top quark acquires its mass 
\BE
m_t \sim  \frac{\left< \overline{U_L} U_R \right> }{\Lambda^2}.
\label{inmt}
\EE
Because the condensate also gives mass to the $W$ boson,
one may naively expect that the condensate has a same order of magnitude
as the electroweak symmetry breaking scale,
\BE
\left<  \overline{U_L} U_R \right>^{1/3} \sim \Lambda_{EW} 
\sim \mbox{1~TeV}.
\label{invev}
\EE
From the observed value of the top-quark mass 
$m_t \simeq 175 ~ \GeV$\cite{mt}, 
this naive argument suggests that
$1/\Lambda^2 \sim m_t/\Lambda^3_{EW}$, and that
the new physics scale
$\Lambda$ is not too far from $\Lambda_{EW}$.

During the last decade,
there were many analyses of the 
dynamical symmetry breaking scenarios in the case of
a large top-quark mass $m_t > 100$~GeV.
In 1985, Appelquist, et al.\cite{app} 
studied the $\rho$ parameter ($T$ parameter)
in the context of
extended technicolor models.
They pointed out that naively 
the mass difference between the techni-$U$ and 
its iso-partner techni-$D$
is proportional to the top-quark mass, so that this difference would
contribute to the $T$ parameter.
Also, they noted that an extra isospin violating operator
\BE
\frac{1}{\Lambda'^2}
\overline{Q_R} \gamma^{\mu} \sigma^3 Q_R
\overline{Q_R} \gamma_{\mu} \sigma^3 Q_R \label{ib4f} ,
\EE
where $Q_R = (U_R, D_R)^T$,
may give a large contribution to the $T$ parameter
since $\Lambda'$ is considered to approximate $\Lambda$.
(For $m_t \simeq 175$~GeV, the latter effect would be more significant
than the former.)
It was suggested in Ref.\cite{Chivukula} 
that the $T$ parameter would be enhanced
in the walking technicolor scenario.
More detailed analyses on the $T$ parameter were given later in
Refs.\cite{atew,en}.
Recently the experimental constraint on the $T$ parameter
has become more
severe, and deviation from the standard model
prediction is seen to be very small\cite{Matsumoto}.
Reflecting the present constraint, some dynamical symmetry breaking
models have been proposed\cite{holdommodel,asakamodel}
in which the operator (\ref{ib4f})
is suppressed at low energy.
Ref.\cite{cdt} studied the constraints from 
the present $T$ parameter and top mass data,
and discussed the top-color assisted technicolor
model in this context.

In this paper we assume that at low energy the four fermi 
operators other than (\ref{in4fint})
can be neglected.
We estimate the strength and the intrinsic mass scale of the
particular operator (\ref{in4fint}) in detail on this assumption.
We use unitarity argument to quantify the strength of the operator.
Then we estimate the contribution of this operator
to the $T$ parameter.
We include the QCD correction to the top-quark mass function and
study its effect on the $T$ parameter.

In order to incorporate the dynamics of symmetry breaking into our
analyses, 
we solve numerically
the Schwinger-Dyson and Bethe-Salpeter equations in the improved-ladder
approximation\cite{improved-ladder}. 
We follow the formalism developed in Refs.\cite{Kugo_FPI,meson,Kugo4};
In these papers,
taking $f_\pi = 94$~MeV as an only input parameter for QCD, the
quantities 
$\Lambda_{QCD}$, $\left< \overline{\Psi} \Psi \right>$,
$m_\rho$, $m_{a_1}$, $m_{a_0}$, $f_\rho$ and $f_{a_1}$ have been
calculated, which meet the experimental values within 
20$\sim$30\% accuracy for $\Lambda_{QCD}$,$\cdots$,$m_{a_0}$ and within
a factor of 2 for $f_\rho$ and $f_{a_1}$.
Thus, we expect to study the dynamical effect semi-quantitatively using
the formalism.

In Section 2 we present our assumption on the dynamical symmetry
breaking scenario.
Then we estimate the strength of the four fermi operator 
from the observed top-quark mass and $W$ boson mass
in Section 3.
Using the result, 
the contributions to the $T$ parameter are estimated in Section 4.
Conclusion and discussion are given in Section 5.

The explicit formulas of the Schwinger-Dyson and Bethe-Salpeter
equations, as well as other equations used in our numerical analyses,
are collected in Appendix.
\section{Symmetry Breaking Sector and Four Fermi Operator}

In this section
we explain our assumption on the scenario of 
dynamical generation of the top-quark mass. 

First, we assume technicolor-like scenario\cite{Weinberg-Susskind}
for breaking electroweak gauge symmetry.
We introduce non-standard-model fermions 
following the one-doublet technicolor (TC) model as
\BE
Q_L = {U \choose D}_L, \qquad U_R, \qquad D_R 
\label{tf} .
\EE
The weak hypercharges are assigned as $Y(Q_L)=0$, $Y(U_R) = 1/2$, and 
$Y(D_R) = -1/2$.
These fermions belong to the fundamental representation
of $SU(N_{TC})$ gauge group, and 
they form the pair condensates
\BE
\left< \overline{U_L} U_R \right> \neq 0
\qquad  \mbox{and} \qquad
\left< \overline{D_L} D_R \right> \neq 0 \label{condensate}
\EE
via the $SU(N_{TC})$ gauge interaction.
Later when we solve
the Schwinger-Dyson equations numerically, 
we will deal with both the
technicolor-like and walking-technicolor-like\cite{WTC} scenarios
by varying the running behavior of the gauge coupling constant.
In the following analyses,
we consider only the cases $N_{TC} = 2$ and 3
taking into account the present stringent experimental
constraint\cite{Matsumoto} on the $S$ parameter\cite{STU}.

Secondly, in order to generate the top-quark mass, 
we introduce an effective four fermi operator
\BE
\frac{G}{M^2} ~ \left( \overline{Q_L} U_R \right) 
\left( \overline{t_R} q_L \right) ~ + h.c., \label{4fop}
\EE
where $q_L$ denotes the ordinary quark doublet $(t_L,b_L)^{T}$.
$G$ is a dimensionless coupling and $M$ is the intrinsic mass
scale of this operator.
Because the four fermi operator cannot be a fundamental interaction,
there should be some energy scale above which this operator 
will resolve, and we call this scale $M$.
In other words, it is the scale where higher dimensional operators
neglected in Eq.(\ref{4fop}) become relevant.
We may neglect the higher dimensional operators if
all the energy scales involved in  our calculation
satisfy $E/M \ll 1$.
In particular, since
we will incorporate the non-perturbative dynamics of 
the $SU(N_{TC})$ technicolor interaction by solving the Schwinger-Dyson 
and Bethe-Salpeter equations, the validity of our 
effective treatment of the four fermi operator (\ref{4fop})
as a contact interaction would be justified if
the technicolor scale $\Lambda_{TC}$ 
satisfies $\Lambda_{TC} \ll M$.

We assume that we may 
neglect all effective four fermi operators
other than (\ref{4fop}) which would be induced at low energy
in the models such as extended technicolor models\cite{ETC}.
(See, however, discussion in Section 5.)
This is because an operator such as Eq.(\ref{ib4f}) would give a very
large contribution to the $T$ parameter.
We do not consider the dynamical origin of the operator (\ref{4fop})
in this paper.
%
%
%
\section{Strength and Mass Scale of Four Fermi Operator}

In this section, we estimate the strength $G$ and 
the intrinsic mass scale $M$ of the
four fermi operator (\ref{4fop}) 
from the observed top-quark and $W$ boson masses.

\subsection{Relation between $G$ and $M$}

First, we solve numerically the Schwinger-Dyson equation 
depicted diagrammatically in Fig.\ref{fig-sdeq} for 
the mass function $\Sigma$
of techni-$U$ or techni-$D$ fermion\cite{Kugo_FPI}.
In order to set mass scale in the numerical calculation,
we use the charged decay constant 
$F_{\pi^\pm}$, which is obtained by solving the homogeneous
Bethe-Salpeter equation\cite{Kugo_FPI} 
or using the Pagels-Stokar's formula\cite{Pagels-Stokar1}.
(Both results are in good agreement.)
From the $W$ boson mass $M_W$, 
$F_{\pi^\pm}$ is normalized as
\BE
F_{\pi^\pm}  =  \frac{2 M_W}{g} \simeq 250~\mbox{GeV},
\label{mass-scale}
\EE
where $g$ is the $SU(2)_L$ gauge coupling constant.

As shown in Fig.\ref{fig-tmass}, the
top quark acquires its mass $m_t$
through the four fermi operator (\ref{4fop}), and 
$m_t$ can be calculated as
\BE
m_t = \frac{G}{M^2} ~ \left< \overline{U_L} U_R \right>_M,
\label{topmass} 
\EE
where
\BE
\left< \overline{U_L}U_R \right>_M &=& 
\frac{1}{2}
		 \int_{p_E^2 \leq M^2}
                 \frac{d^4 p}{( 2 \pi)^4} 
		\mbox{tr} 
		\left( \frac{i}{ \not\!p - \Sigma(p) } \right)
\nonumber
\\
	&=& \frac{ N_{TC} }{ 8 \pi^2}
		\int^{M^2}_0 dx 
		\frac{ x \Sigma(x) }{ x + \Sigma(x)^2 },
\label{vev}
\EE
with $x=p_E^2 =-p^2$.
Note that we define the intrinsic mass scale $M$ of the 
four fermi operator 
(\ref{4fop}) as the momentum cut-off of the integral
in Eq.(\ref{vev})
since the four fermi operator will resolve
above the energy scale $M$.

By calculating the condensate $\left< \overline{U_L} U_R \right>_M$ 
for a given $M$, and substituting the top-quark mass 
$m_t \simeq 175 ~ \GeV$\cite{mt} 
in Eq.(\ref{topmass}), we obtain the coupling $G$ as a function of $M$. 
We show the $G$-$M$ relation in Figs.\ref{fig-gm}
for the $SU(2)$ and $SU(3)$ technicolor cases,
and also for the walking technicolor case.%
\footnote{
In our analyses, the walking technicolor case corresponds to 
the $SU(3)$ technicolor theory with one technifermion doublet
which is introduced in (\ref{tf}) and
ten technifermion singlets under $SU(2)_L\times U(1)_Y$.
The 1-loop $\beta$-function reduces to approximately 1/3 of the 
$SU(3)$ technicolor case.}
We neglect the region $M \lsim \Lambda_{TC}$
where our effective treatment of the four fermi operator (\ref{4fop})
as a contact interaction breaks down.
We define $\Lambda_{TC}$ as the scale where
the leading-logarithmic running coupling constant
of technicolor diverges.
The values of $\Lambda_{TC}$ in our numerical estimates 
are
$\Lambda_{TC}$ $\simeq$ 1.7, 1.3, and 0.6 TeV
for the $SU(2)$, $SU(3)$ technicolor, and the walking technicolor
cases, respectively.
For the technicolor cases, $G(M)$ is almost proportional
to $M^2$ as $M$ increases, while it is almost proportional to
$M$ for the walking technicolor case.
These tendencies are consistent with the 
asymptotic behaviour of the mass function 
of techni-$U$\cite{polizer,Kugo3}:
\BE
\Sigma (x) \sim \frac{1}{x} (\log x)^{\frac{3C_2}{\beta_0}-1},
\EE
where $\beta_0$ is the 1-loop $\beta$ function of the technicolor
interaction and $C_2 = (N_{TC}^2 - 1)/2N_{TC}$.
It should be noted that for both technicolor and walking technicolor 
cases, the coupling $G$ should be rather strong,
typically $G/4\pi \sim \order(1)$ in order to explain the 
observed top-quark mass.
%
%
%
%
\subsection{Unitarity Constraint for the Coupling $G$}

We have seen that 
the coupling $G$ should 
be quite large.
As a criterion for testing the strength of $G$, we study 
tree-level unitarity limit related to the four fermi operator
(\ref{4fop}). 
There are a few scattering amplitudes 
induced by this operator at tree-level
which increase in high energy and 
at some energy would violate the unitarity bound.
The tree-level unitarity violation occurs at lower energy
for larger value of $G$ in general.
However, the energy to reach the unitarity limit
should be above the scale  
$M$, since we have assumed that the four fermi operator (\ref{4fop}) can 
be treated as a contact interaction below the scale $M$, that is,
the higher dimensional operators
are irrelevant at energy scale $E \ll M$.
We see that this requirement leads to the upper bound for $G$.

Let us consider the two-body to two-body scatterings of fermions
via the operator (\ref{4fop}) at the energy scale where
$E \gg \Lambda_{TC}$.
In this energy region  the confinement effect of technicolor
may be ignored.
The tree-level matrix elements of these processes increase quadratically 
with the center of mass energy.
We find that the scattering amplitude for
$t\overline{t} \rightarrow U\overline{U}$ in $J=0$ channel
will reach the unitarity limit most quickly.
The partial-wave amplitude is given by
\begin{eqnarray}
T^{J=0}\left(\sqrt{s}\right) = \frac{ \sqrt{N_C N_{TC}} }{8 \pi}
\frac{ G }{ M^2 } ~ s,
\label{matrix-element}
\end{eqnarray}
where $\sqrt{s}$ is the center of mass energy.%
\footnote{In our previous paper\cite{Asaka}, we incorrectly omitted
the color and technicolor factors in Eq.(\ref{matrix-element}) 
which come from the normalization of the initial and final states.
}\
We set $m_t = m_U = 0$ considering $E \gg \Lambda_{TC}$.
Unitarity limit\cite{iz} for an inelastic scattering channel is given by 
$\left| T^J \right| \leq 1$.
We may demand that the tree-level unitarity should not be violated below
$\sqrt{s} = M$, that is,
\begin{eqnarray}
\left| T^{J=0}(\sqrt{s} = M) \right| \leq 1,
\end{eqnarray}
which can be translated to the upper bound for $G$ as
\begin{eqnarray}
G \leq \frac{ 8 \pi }{ \sqrt{ N_C N_{TC} } }.
\end{eqnarray}

The bound is so stringent that there are hardly allowed regions in 
the $G$-$M$ planes in Figs.~\ref{fig-gm} for $M >\Lambda_{TC}$. 
This result suggests that our effective treatment of 
the four fermi operator as a contact interaction 
breaks down.
Since the coupling $G$ exceeds the perturbative unitarity limit
for $M > \Lambda_{TC}$,
the higher order corrections of $G$ are large and
should modify the tree level amplitude to restore unitrairy
at energy scale $E \sim \Lambda_{TC}$.
Such corrections induce the higher dimensional operators
which become relevant at energy scale $E \sim \Lambda_{TC}$.
Thus the scale $M$ above which the operator (\ref{4fop}) will resolve
is found to be around $\Lambda_{TC}$ or less.

\subsection{Coupled Schwinger-Dyson Equations}
From the above discussion, the coupling $G$ should be strong 
and the non-perturbative effect of the
four fermi operator (\ref{4fop}) would be significant.
Meanwhile, in Subsection 3.1, 
we only considered the \order($G$) contribution 
in estimating the top-quark mass.
Here we include part of the non-perturbative effect of the
four fermi operator
and re-estimate the $G$-$M$ relation.

For this purpose,
we solve the coupled Schwinger-Dyson equations 
for techni-$U$ and top quark
shown diagramatically in Fig.\ref{fig-csdeq}.
Note that the operator (\ref{4fop}) affects 
the mass function of techni-$U$ but not that of techni-$D$.
We re-estimate the coupling $G(M)$, and
the results are shown in Figs.\ref{fig-gmcsd}.
$G(M)$ is found to decrease compared
to the previous analysis.
This can be understood by noting that the top-quark loop
diagram in Fig.\ref{fig-csdeq} gives additive contribution to 
$\left< \overline{U_L} U_R \right>_M$ so that a
weaker coupling $G$ is
necessary to generate the top-quark mass.
The deviations from the previous analyses increase for larger $M$ 
since the coupling $G$ is larger in this region and
the non-perturbative effect is more significant.

\subsection{QCD correction}

Finally we incorporate QCD correction in estimating the $G$-$M$
relation. 
In the previous analyses, we neglected QCD running 
effect of the top-quark mass function $\Sigma_t$.
From the renormalization group equation 
analysis, $\Sigma_t$ receives QCD correction
from $m_t$ to $M$ scale as
\BE
\Sigma_t(M^2) =
\Sigma_t({m_t}^2) 
\left[
\frac{\log( {m_t}^2 /\Lambda_{QCD}^2)}
     {\log(     M^2 /\Lambda_{QCD}^2)}
\right]^{\frac{\scriptstyle 4}{\scriptstyle \beta}},
\label{rge-ope}
\EE
where $\beta=11-\frac{2}{3}n_f$ 
is the lowest order coefficient of the $\beta$-function of
renormalization group equation of QCD.

We include this running effect by solving
the coupled Schwinger-Dyson equations which are shown 
in Fig.\ref{fig-csdqcd}.
(See Appendix for detail.)
Again, the coupling $G(M)$ is obtained.
The results are given in Figs.\ref{fig-gmqcd}.
We find that the coupling becomes smaller in each case.
This is because for $\mu > m_t$ the top-quark mass becomes smaller,
$\Sigma_t(\mu^2)<m_t$, 
due to the QCD correction Eq.(\ref{rge-ope}).
Nevertheless there are still no allowed regions in the $G$-$M$ planes
for $M > \Lambda_{TC}$.
%
%
\section{Contributions to $T$ parameter}

Because the four fermi operator (\ref{4fop}) violates 
isospin symmetry, one may expect that the results obtained in the
previous section may lead to large isospin violating effects.
In this section we estimate the contributions of the four fermi operator
to the $T$ parameter.

\subsection{Contribution of $\Sigma_U - \Sigma_D$}
Here we consider the isospin violating effect originating from the
difference of techni-$U$ and techni-$D$ mass functions.
The contribution of the
four fermi operator to the techni-$U$ mass function
in the Schwinger-Dyson equations causes this difference.
(See Fig.\ref{fig-csdeq}.)
We estimate the $T$ parameter 
and compare with the present experimental bound,
from which we
extract another bound for the mass scale $M$ of the four fermi operator.

We calculate the charged and neutral decay constants
$F_{\pi^\pm}$ and $F_{\pi^0}$ from the mass functions of
techni-$U$ and $D$ using the generalized Pagels-Stokar's
formula\cite{Pagels-Stokar2}.
Then the contribution to the $T$ parameter ($T_{NEW}$)
is calculated as
\BE
\alpha T_{NEW}
=
\frac{ F_{\pi^\pm}^2 - F_{\pi^0}^2 }{F_{\pi^0}^2},
\EE
where $\alpha$ = 1/128 is the fine structure constant.
Thus, we can calculate $T_{NEW}$ as a function
of $G$ and $M$.

Let us first consider the case discussed in Subsection 3.3,
that is, we neglect the QCD effect on $\Sigma_t$.
The results are shown in
Figs.\ref{fig-Tparm}
when the coupling $G$ is on the corresponding lines 
$G = G(M)$ in Fig.\ref{fig-gmcsd}.
One sees that $T_{NEW}$ increases with $M$ (or $G$).
This behavior is consistent with the naive
estimate of $T$ parameter by the fermion
1-loop calculation\cite{STU}
\BE
T \simeq 
\frac{N_{TC}}{12 \pi \sin^2 \theta_W \cos^2 \theta_W}
\left[ \frac{ ( \Delta m )^2 }{ M_Z^2 } \right],
\label{naiveest1}
\EE
combined with a naive estimate of the mass difference
of techni-$U$ and $D$ from the coupled Schwinger-Dyson
equations (i.e. the additional term in Fig.\ref{fig-csdeq})
\BE
\Delta m \simeq \frac{N_C}{8 \pi^2} \, G(M) \, m_t .
\label{naiveest2}
\EE
Also we see that $T_{NEW}$ is larger for
the walking technicolor case than that of technicolor case
for the same $M$.
This tendency has been pointed out by Chivukula\cite{Chivukula}.

Comparing the results with the present experimental 
data on the $T$ parameter\cite{Matsumoto}
\BE
T_{exp} - T_{SM}( m_t = 175~ \GeV, m_H = 1~ \TeV ) = 0.32 \pm 0.20,
\EE
we may put 3$\sigma$ upper bounds for the intrinsic mass
scale $M$ as follows:
\BE
&M \lsim 7~ \TeV& \qquad 
\mbox{ for $SU(2)$ technicolor } \nonumber \\
&M \lsim 5~  \TeV& \qquad 
\mbox{ for $SU(3)$ technicolor } \nonumber \\
&M \lsim 4~  \TeV& \qquad 
\mbox{ for walking technicolor }
\EE
Note that the bound is more stringent for the walking 
technicolor case.

Next, we inculde the QCD correction on $\Sigma_t$.
The results are also shown in Figs.\ref{fig-Tparm}.
Note that the QCD correction reduces $T_{NEW}$ by about 40\%.
This can be understood from Eqs.(\ref{naiveest1}) and (\ref{naiveest2})
if we note that both $G(M)$ and $\Sigma_t(\mu^2)$ ($\mu>m_t$)
get smaller by the QCD correction.
(See Subsection 3.4.) \
Similarly, 3$\sigma$ upper bounds for $M$ are obtained:
\BE
&M \lsim 12~ \TeV& \qquad 
\mbox{ for $SU(2)$ technicolor } \nonumber \\
&M \lsim 9~  \TeV& \qquad 
\mbox{ for $SU(3)$ technicolor } \nonumber \\
&M \lsim 4~  \TeV& \qquad 
\mbox{ for walking technicolor }
\label{Tqcd}
\EE

\subsection{Contribution of 
$\overline{U_R} \gamma_{\mu} U_R \overline{U_R} \gamma^{\mu} U_R$}

We started our analyses assuming that only the four fermi operator
(\ref{4fop}) exists at low energy in order to dispense with the
potentially dangerous operator
\BE
C \, \overline{U_R} \gamma_{\mu} U_R \overline{U_R} \gamma^{\mu} U_R ,
\label{D4fop}
\EE
which would induce a large $T$ parameter\cite{app}.
We found in the previous section, however, that the higher order
corrections of the operator (\ref{4fop}) cannot be neglected.
In fact, the above operator (\ref{D4fop})
is generated by four insertions of the operator 
(\ref{4fop}) at three-loop level.\
(Fig.\ref{fig-uuuu})\
From a dimensional analysis
of this graph, we estimate
\BE
C 
&\sim& - \frac{N_C^2}{(4\pi)^6} \frac{G^4}{M^2}.
\label{codop}
\EE
Then the contribution of the operator (\ref{D4fop})
to the $T$ parameter can be estimated as
\BE
T \sim 6 \times 10^{-2}\, \frac{N_{TC}(N_{TC}+1)}{12}
\left(\frac{m_U}{\mbox{1$\,$TeV}} \right)^4 
\left(\frac{\mbox{2$\,$TeV}}{M}\right)
^2 \left(\frac{G}{4\pi}\right)^4 
\left( \log \frac{\Lambda_{TC}^2}{m_U^2} \right)^2 .
\label{daot}
\EE

We should note that the three-loop graph is very sensitive to the
cut-off of the loop momenta. 
Therefore the estimated value Eq.(\ref{daot}) may change by a factor 
$\sim 10$ by a slight change of the cut-off and therefore it may give
a non-negligible contribution to the $T$ parameter.
We should also remark that Eq.(\ref{codop})
may suggest self-inconsistency of our assumption 
that we neglect all four fermi operators other than (\ref{4fop}).
We will discuss this point in the next section.

\section{Conclusion and Discussion}

In this paper, within a scenario where the top-quark mass is
generated dynamically, we estimated the coupling $G$ and the intrinsic
mass scale $M$ of the four fermi operator that induces the top-quark
mass. 
Also, we studied the contribution of this four fermi operator to the $T$ 
parameter.

Throughout our analyses, we made the following assumptions:
\begin{itemize}
\item
The $W$ and $Z$ bosons acquire their masses in the one-doublet 
technicolor-like scenario.
\item
The top quark acquires its mass via the effective four fermi operator
(\ref{4fop}). 
We consider only this four fermi operator and neglect all other 
effective four fermi operators that may be induced in various
dynamical models. 
\end{itemize}
We 
incorporated the dynamics of $SU(N_{TC})$ gauge interaction by solving
the Schwinger-Dyson and Bethe-Salpeter equations numerically in the
improved-ladder approximation
(in all the analyses except in Subsection 4.2).

In Section 3, we studied in detail the strength $G$ and the
intrinsic mass scale $M$ of the four fermi operator using 
$M_W$ and $m_t$ as the input parameters.
We obtained $G$ as a function of $M$ in the region  
$M > \Lambda_{TC}$, and found that $G$ is rather strong,
$G/4\pi \sim O(1)$.
Then we compared the coupling $G$ with that demanded by
the tree-level unitarity bound.
Our results suggest that $M$ should be of the order of
$\Lambda_{TC} \simeq 1 \sim 2$~TeV or less, so that 
the four fermi operator cannot be treated as `point-like'
at scale $E \sim \Lambda_{TC}$.
Conventionally the four fermi operator (\ref{4fop}) has been treated 
perturbatively in many papers, but the unitarity saturation shows that
such a treatment is inconsistent with the presently observed 
top-quark mass.
We included part of the higher order corrections of the four fermi
operator (\ref{4fop}) by solving the coupled Schwinger-Dyson equations.
Also we included the effect of QCD correction on the top-quark mass
function. 
These effects, respectively, are found
to reduce $G(M)$.

In Section 4 we studied the contributions of the four fermi operator 
(\ref{4fop}) to
the $T$ parameter.
First we estimated the contribution of the difference between the mass
functions of techni-$U$ and techni-$D$.
We found that the QCD correction is large and reduces the
contribution to the $T$ parameter by about 40\%.
The estimated $T$ parameter
is within the present experimental bound.
Then we used the experimental bound 
to obtain another upper bound for $M$, and found that typically 
$M$ is less than 10 TeV.
The bound on $M$ is more stringent for the walking technicolor case.
Secondly we pointed out that the dangerous operator
$\overline{U_R} \gamma_{\mu} U_R \overline{U_R} \gamma^{\mu} U_R$
would be generated by the four fermi operator (\ref{4fop}) 
at the three-loop level, and estimated its contribution to the $T$
parameter from a dimensional analysis.
The contribution may become non-neglegible.

We found that the four fermi operator (\ref{4fop}) cannot be
treated as `point-like' at scale $E \sim \Lambda_{TC}$. 
In order to make a more consistent analysis, one needs to
specify the `structure' of the four fermi operator, i.e.\ specify the 
dynamical origin of this operator.
One way is to rewrite the four fermi operator in terms of a 
massive-gauge-boson exchange interaction 
as in the extended technicolor models.
We are currently making further analyses in this direction.

We started our analyses on the assumption that all four fermi 
operators except Eq.(\ref{4fop}) can be neglected.
We found, however, that other four
fermi operators generated in higher orders of the operator (\ref{4fop})
may be non-negligible.
(e.g.\ The operator 
$\overline{U_R} \gamma_{\mu} U_R \overline{U_R} \gamma^{\mu} U_R$.)
This self-inconsistency seems to put certain constraints when constructing
a viable model of dynamical electroweak symmetry breaking.
Namely, suppose one could construct an extended technicolor model that 
has ETC gauge bosons which induce only the 
four fermi operator (\ref{4fop}) at tree level.
Then 
other four fermi operators induced at higher loops would be suppressed
by powers of $\Lambda_{TC}/M$, but this factor is 
close to one for the top-quark mass $\simeq 175$~GeV.

\section*{Acknowledgments}

We are grateful to K.~Fujii, K.~Hagiwara,
K.~Hikasa, J.~Hisano, B.~Holdom,
N.~Maekawa, T.~Moroi, H.~Murayama, M.~Peskin,
and J.~Terning
for fruitful discussion.

\newpage

\newcommand{\dirac}{\not \hspace{-0.9mm}}
\newcommand{\bea}{\begin{eqnarray}}
\newcommand{\eea}{\end{eqnarray}}
\newcommand{\simgt}{\hbox{ \raise3pt\hbox to 0pt{$>$}\raise-3pt\hbox{$\sim$} }}
\newcommand{\simlt}{\hbox{ \raise3pt\hbox to 0pt{$<$}\raise-3pt\hbox{$\sim$} }}

\section*{Appendix}
In this appendix, 
we list the Schwinger-Dyson equations as well as other formulas
which are used in our numerical analyses.
In Section 3, we solved the coupled and non-coupled
Schwinger-Dyson equations in the improved ladder approximation
for the mass functions of techni-$U$, techni-$D$ and top quark
($\Sigma_U$, $\Sigma_D$ and $\Sigma_t$).
All these equations can be written in the following forms:
\bea
&&
\Sigma_U(x) = 
\frac{\lambda(x)}{4x} \int_0^x dy 
\frac{y\Sigma_U(y)} {y+\Sigma_U^2(y)} 
+ 
\int_x^{\Lambda^2} dy 
\frac{\lambda(y)\Sigma_U(y)} {4 \left( y + \Sigma_U^2(y) \right)}
\nonumber \\
&&~~~~~+
A_1 \cdot \frac{N_C}{ 8 \pi^2} \frac{ G }{M^2}
\int_0^{M^2} dy \frac{y \Sigma_t(y)}{y+\Sigma_t^2(y)}, 
\nonumber \\
&&
\Sigma_D(x) =
\frac{\lambda(x)}{4x} \int_0^x  dy \frac{y \Sigma_D(y)}
{y + \Sigma_D^2(y)}
+ 
\int_x^{\Lambda^2} dy \frac{ \lambda(y) \Sigma_D(y)}
{ 4 \left( y + \Sigma_D^2(y) \right)}, 
\nonumber \\
&&\Sigma_t(x) =
\frac{N_{TC}}{8 \pi^2} \frac{G}{M^2}
\int_0^{M^2} dy
\frac{y \Sigma_U(y)}{y+\Sigma_U^2(y)} 
\nonumber \\
\nonumber \\
&&~~~~+
A_2 \left[  
\frac{\lambda_{QCD}(x)}{4x} \int_0^x  dy \frac{y \Sigma_t(y) }
{y+\Sigma_t^2(y)}
+
\int_x^{\Lambda^2} dy \frac{\lambda_{QCD}(y) \Sigma_t(y)}
{4 ( y + \Sigma_t^2(y) )}
\right],
\label{ap-allsdeq}
\eea
\\
where $\lambda(x)$ and $\lambda_{QCD}(x)$ denote the running coupling
constants for technicolor and color interactions, respectively.

According to Ref.\cite{Kugo_FPI}, we take $\lambda(x)$ 
as follows:
\bea
\lambda(x) &=& \lambda_0 \times 
\left\{ 
\begin{array}{ll}
C 
&
\mbox{if $t \leq t_0$}\\
C - 
\displaystyle{ 
 \frac{1}{2} \frac{A}{(1+A t_{IF})^2 }
\frac{(t-t_0)^2}{ ( t_{IF} - t_0 ) }
}
&
\mbox{if $t_0 \leq t \leq t_{IF}$}\\
\displaystyle{
\frac{1}{ 1 + At}
}
&
\mbox{if $t_{IF} \leq t$}
\end{array} 
\right. 
\label{defcoup}
\end{eqnarray}
with
\begin{eqnarray}
t = \ln x~~~\mbox{and}~~~ 
C =   \frac{1}{2} 
\frac{ \displaystyle{ A( t_{IF} - t_0 )} }{(1+A t_{IF})^2 } 
   +  \frac{1}{ 1 + A t_{IF} },
\nonumber
\eea
%
where $\lambda_0/A = 12 C_2/\beta_0$ and $\beta_0$ is the
1-loop order coefficient of the $\beta$-function and
$C_2=(N_{TC}^2-1)/2N_{TC}$ represents the second Casimir.    
Thus, above the infrared cut-off scale
$t \ge t_{IF}$, $\lambda(x)$ is related to the 
1-loop running coupling constant $g_{TC}(x)$ as
\bea
\lambda(x) &=&
\frac{3}{4\pi^2} ~ C_2 ~ g_{TC}^2(x).
\eea
In our numerical calculation, we fix the point $t_0 = \ln \mu^2_0$ 
relative to $\Lambda_{TC}$ and consider the infrared cut-off
scale $t_{IF}$ as a free parameter.
We define $\Lambda_{TC}$ as the point where
the leading logarthmic running coupling constant diverges:
\bea
 1 + A \ln \Lambda_{TC}^2=0.
\eea
As for all the dimensionful quantities in our calculation, we
set scale by normalizing the decay constant as in
Eq.(\ref{mass-scale}).

For the QCD coupling constant, $\lambda_{QCD}(x)$ takes the same form 
as Eq.(\ref{defcoup}). 
Above the infrared cut-off scale
of QCD,
$\lambda_{QCD}(x)$ can be expressed by 
the 1-loop running coupling constant $g_{QCD}(x)$ as
\bea
\lambda(x) &=&
\frac{3}{4\pi^2} ~ C_2^{QCD} ~ g_{QCD}^2(x).
\eea
where $C_2^{QCD}=(N_{C}^2-1)/2N_{C}$ and $N_{C}=3$. 
We set the mass scale of QCD by taking
$\Lambda_{QCD}$ = 200 MeV.
\begin{enumerate}
\item
In Subsection 3.1, we solved the equation (\ref{ap-allsdeq}) 
for the techni-$U$ setting
\bea
A_1 = 0,~~~\mbox{and}~~~~\Lambda = \infty,
\label{ap-sdeq}
\eea
which is given diagrammatically in Fig.\ref{fig-sdeq}.
\item
In Subsection 3.3, we solved the coupled Schwinger-Dyson equations 
(Fig.\ref{fig-csdeq}), which correspond to
\bea
A_1 = 1 ,~~~A_2 = 0, ~~~ \mbox{and} ~~~\Lambda = M.
\label{ap-csdeq}
\eea
\item
In Subsection 3.4, we solved the coupled Schwinger-Dyson equations
including the QCD correction (Fig.\ref{fig-csdqcd}), 
which correspond to
\bea
A_1 = 1 ,~~~A_2 = 1, ~~~ \mbox{and}~~~ \Lambda = M.
\label{ap-csdqcdeq}
\eea
\end{enumerate}
For the above coupled Schwinger-Dyson equations (\ref{ap-csdeq})
and (\ref{ap-csdqcdeq}), we take the ultraviolet cutoff
scale\cite{cutoff} 
as $\Lambda =M$.

As mentioned earlier,
we calculated the charged decay constant 
$F_{\pi^\pm}$ in order to set the mass scale.
We define the decay constant as,
\bea
\left<0 \right|
\bar{Q}_L \gamma^\mu T^b Q_L (0) \left|\pi^a(q)\right>
=\frac{i}{2} F^{a b}_{\pi}q^\mu,
\eea
where $T^a$ ($a$=1,2,3) is the generator of $SU(2)_L$.
Then the charged and neutral decay constants, respectively, are given by 
$F_{\pi^\pm} \equiv F_{\pi}^{11} = F_{\pi}^{22}$ and 
$F_{\pi^0} \equiv F_{\pi}^{33}$.

For the isospin symmetric case ($F_{\pi^0} = F_{\pi^\pm} = F_\pi$),
we obtained $F_\pi$ by solving the homogeneous Bethe-Salpeter (BS)
equation\cite{Kugo_FPI},
or using the Pagels-Stokar's formula\cite{Pagels-Stokar1}.

According to Ref.\cite{Kugo_FPI}, 
the BS amplitude $\chi$ for the $J^{PC}=0^{-+}$
massless state (the techni-pion state) is defined by
\bea
&&\int d^4 r ~ e^{ipr} 
\left< 0 \right| T ~
\Psi_\alpha^{f,i}( x + r/2 ) \overline{\Psi}_\beta^{f',j} ( x - r/2 )
\left| \pi^a(q) \right>
\nonumber \\
&&~~~~~~~~~~~~~~~~~~~
= 
\frac{1}{\cal N} ~ \delta_{ij} ~ T^a_{ff'} ~ e^{-iqx} ~
\chi_{\alpha \beta}(p,q)
\eea
where $i,j,\cdots$ denote the technicolor indices, 
$f, f',\cdots$ the flavor indices, and $\alpha,\beta,\cdots$ 
the spinor indices, and $\Psi$ denotes the techni-fermion.
$\cal N$ is introduced as the normalization for the amplitude.
From the spinor structure and the quantum number $J^{PC} = 0^{-+}$,
the bispinor part $\chi_{\alpha \beta}$ is expanded into following
invariant amplitudes:
\bea
\chi(p,q) = \left[ S(p;q) + P(p;q) (p \cdot q) \not{p} 
+ Q(p;q) \not{q} + \frac{1}{2} T(p;q)  
\left( \not{p} \not{q} - \not{q}\not{p} \right)
\right] \gamma_5.
\nonumber\\
\label{dummy}
\eea
Here the above amplitudes are found to be even functions
of $(p \cdot q)$ from the charge conjugation property.
Using the on-shell condition of $\pi$ ($q^2=0$),
each amplitude can be expanded as
\bea
S(p,q) &=& S_0(p^2) + \sum_{I=1}^{\infty} (p \cdot q)^{2I} S_I(p^2),
\nonumber \\
P(p,q) &=& P_0(p^2) + \sum_{I=1}^{\infty} (p \cdot q)^{2I} P_I(p^2), 
~~~ \mbox{etc}. \label{expand-of-chi}
\eea
Then we can write down the decay constant $F_{\pi}$ in
terms of coefficients of above expansion.
The definition of $F_{\pi}$ leads
\bea
F_{\pi} i q_{\mu} = - \frac{ 1}{ 2 \cal N} 
\int \frac{d^4 p}{ ( 2\pi )^4 } ~
\mbox{tr} \left[ 
\gamma_{\mu} \gamma_5 \chi_{\alpha \beta}(p,q) 
\right],
\eea
and after angular integration, it takes the form
\bea
F_{\pi}^2 = \frac{N_{TC}}{16 \pi^2}
\int_0^\infty dx x \left[ 4 Q_0(x) - x P_0(x)\right],
\label{fpi-hbs}
\eea
where $x = p_E^2$ and we choose ${\cal N} = F_\pi / 2$.
Thus we only need the first terms in the expansion 
Eq.(\ref{expand-of-chi}) to calculate the decay constant.

In order to calculate $P_0(x)$ and $Q_0(x)$, 
we solve the homogeneous Bethe-Salpeter equation for $\chi$
in the improved ladder approximation,
as shown in Fig.\ref{fig-bseq} diagrammatically, as follows:
\bea
&&\left[ \not{p} + \not{q}/2 - \Sigma(p+q/2) \right]
~  \chi(p,q) ~
  \left[ \not{p} - \not{q}/2 - \Sigma(p-q/2) \right]
\nonumber \\
&=& i
\int \frac{d^4k}{ ( 4 \pi)^4 }
\frac{4 \pi^3}{3} \lambda \left( max(p_E^2, k_E^2) \right)
\nonumber \\
&&\times
\gamma^\mu ~  \chi(k,q) ~ \gamma^\nu ~
\frac{1}{ (p-k)^2 }
\left[ g_{\mu \nu} - \frac{ (p-k)_\mu (p-k)_\nu }{(p-k)^2 }
\right],
\eea
where we use $\lambda$ defined in Eq.(\ref{defcoup}).
We expand this BS equation in power of $(p \cdot q)$
and solve it to first order in $(p \cdot q)$ because
$Q_0(x)$ and $P_0(x)$ are first order terms in $q$. 
Explicit forms of this integral equations 
are found in Ref.\cite{Kugo_FPI}.
By using the numerical solutions for them,
we can calculate $F_{\pi}$ using the formula
(\ref{fpi-hbs}).

We also use the Pagels-Stokar's formula%
\cite{Pagels-Stokar1}
to obtain $F_{\pi}$:
\begin{eqnarray}
F_{\pi}^2 = \frac{ N_{TC} }{ 4 \pi^2 }
\int_0^{\infty} dx ~ \frac{x}{4} ~
\frac{ \displaystyle {
4 \Sigma^2- x \frac{d}{dx} \Sigma^2 } }
{(x+\Sigma^2)^2}.
\end{eqnarray}
Both $F_\pi$'s obtained from
the homogeneous Bethe-Salpeter equation
and from the Pagels-Stokar's formula
are in good agreement.

On the other hand,
for the case where the isospin symmetry is broken,
we calculate the charged decay constant $F_{\pi^\pm}$ 
and the neutral one $F_{\pi^0}$ 
using the generalized Pagels-Stokar's formulas\cite{Pagels-Stokar2}: 
\begin{eqnarray}
&F_{\pi^0}^2&
= 
\frac{N_{TC}}{8\pi^2} \int_0^{\infty}  dx ~
 I_0( \Sigma_U, \Sigma_D ), 
\label{ap-ps0}
\\
&F_{\pi^\pm}^2&
= 
\frac{N_{TC}}{8\pi^2} \int_0^{\infty}  dx ~
 I_\pm( \Sigma_U, \Sigma_D ),
\label{ap-pspm}
\end{eqnarray}
with
\begin{eqnarray}
I_0( \Sigma_U, \Sigma_D ) &\equiv&
x ~ 
\frac{
\displaystyle{
\Sigma_U^2 - \frac{x}{4} ~ \frac{d}{dx}\Sigma^2_U
} }
{ (x+\Sigma_U^2)^2 }
+ ~
x ~
\frac{
\displaystyle{
\Sigma_D^2-\frac{x}{4} ~ \frac{d}{dx}\Sigma_D^2
}}{(x+\Sigma_D^2)^2}, 
\\
I_\pm( \Sigma_U, \Sigma_D ) &\equiv&
x ~
 \frac{
\displaystyle{
\Sigma_U^2+\Sigma_D^2-\frac{x}{4}
~ \frac{d}{dx}(\Sigma_U^2+\Sigma_D^2)}
}
{(x+\Sigma_U^2)(x+\Sigma_D^2)}
\nonumber\\
&+&
\frac{x^2}{2} ~ 
\frac{\Sigma_U^2-\Sigma_D^2}{(x+\Sigma_U^2)(x+\Sigma_D^2)}
~ \frac{d}{dx}\log \left[\frac{x+\Sigma_U^2}{x+\Sigma_D^2}\right]
.
\end{eqnarray}
In our calculation, we cut off the integral at $M$
instead of infinity
in Eqs.(\ref{ap-ps0}) and (\ref{ap-pspm}).
The approximation would be valid since
$\Sigma_U(x)$ and $\Sigma_D(x)$ vanishes swiftly
the region $x \gg \Lambda_{TC}^2$.

\clearpage

%
%

\clearpage
\section*{Figure Captions}
\newcounter{Nfig}
\begin{list}{Figure \arabic{Nfig}:}%
{
\usecounter{Nfig}
}
\item{
The graphical representation for the Schwinger-Dyson equation
for the mass function of techni-$U$ (or techni-$D$)
in the improved ladder approximation.
The blob denotes the mass function for techni-$U$.}
\label{fig-sdeq}
\item{
The diagram for the top-quark mass. 
The line with a blob denotes the full propagator for techni-$U$.}
\label{fig-tmass}
\item{
The allowed regions in the $G$-$M$ plane for (A) $SU(N_{TC})$ 
technicolor cases and for (B) walking technicolor case.
Curved lines represent the coupling $G(M)$ 
obtained from Eq.({\protect \ref{topmass}}).
The curves are drawn only in the region $M > \Lambda_{TC}$.
Horizontal lines show the upper bounds for $G$ 
obtained from the unitarity limit.
For $SU(N_{TC})$ technicolor cases, 
the solid and dot-dashed lines correspond to $N_{TC}=2$ and $N_{TC}=3$, 
respectively.
}
\label{fig-gm}
\item{
The graphical representation for the coupled Schwinger-Dyson 
equations for the mass functions of techni-$U$ and top quark.
}
\label{fig-csdeq} 
\item{
The allowed regions in the $G$-$M$ plane obtained by solving
the coupled Schwinger-Dyson equations. 
The notations are same as Fig.\protect{\ref{fig-gm}}.
Previous results 
(Fig.\protect{\ref{fig-gm}}) are
also shown in dots for comparison.}
\label{fig-gmcsd} 
\item{
The graphical representation for the coupled Schwinger-Dyson 
equations for the mass functions of techni-$U$ and top quark
including the QCD correction to the top quark.
}
\label{fig-csdqcd} 
\item{
The allowed regions in the $G$-$M$ plane obtained by solving
the coupled Schwinger-Dyson equations 
including the QCD correction. 
The notations are same as Fig.\protect{\ref{fig-gm}}.
Previous results 
(Figs.\protect{\ref{fig-gm}} and \protect{\ref{fig-gmcsd}}) are
also shown in dots for comparison.}
\label{fig-gmqcd} 
\item{
The contribution to the $T$ parameter, $T_{NEW}$, from 
$\Sigma_U - \Sigma_D$ when the coupling is on the curved
lines $G= G(M)$
in Figs.\protect{\ref{fig-gmcsd}} and \protect{\ref{fig-gmqcd}}, 
for (A) $SU(2)$ technicolor, (B) $SU(3)$ technicolor,
and (C) walking technicolor case.
The dotted lines are $T_{NEW}$
without QCD correction and 
the solid lines are the ones including the QCD correction.
}
\label{fig-Tparm} 
\item{
A three-loop Feynman diagram that induces the operator 
$ \overline{U}_R \gamma^\mu U_R \overline{U}_R \gamma_\mu U_R $.
}
\label{fig-uuuu} 
\item{
The graphical representation for the homogeneous Bethe-Salpeter
equation for the BS amplitude $\chi$.
}
\label{fig-bseq} 
\end{list}
\clearpage
\begin{figure}
\begin{center}
\centerline{ \epsfig{figure=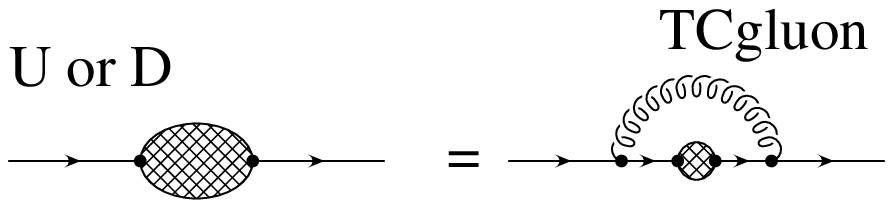,height=3cm} }
\vspace{1cm}
{\Huge Fig. 1}
\end{center}
\end{figure}
\begin{figure}
\begin{center}
\centerline{ \epsfig{figure=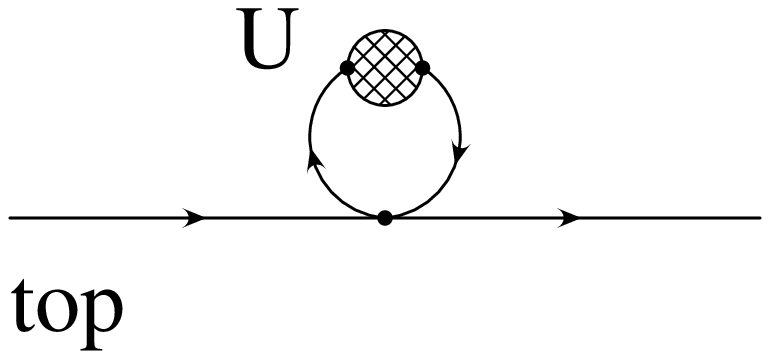,height=4cm} }
\vspace{1cm}
{\Huge Fig. 2}
\end{center}
\end{figure}
\vspace{4cm}
\begin{figure}
\begin{center}
\centerline{ \epsfig{figure=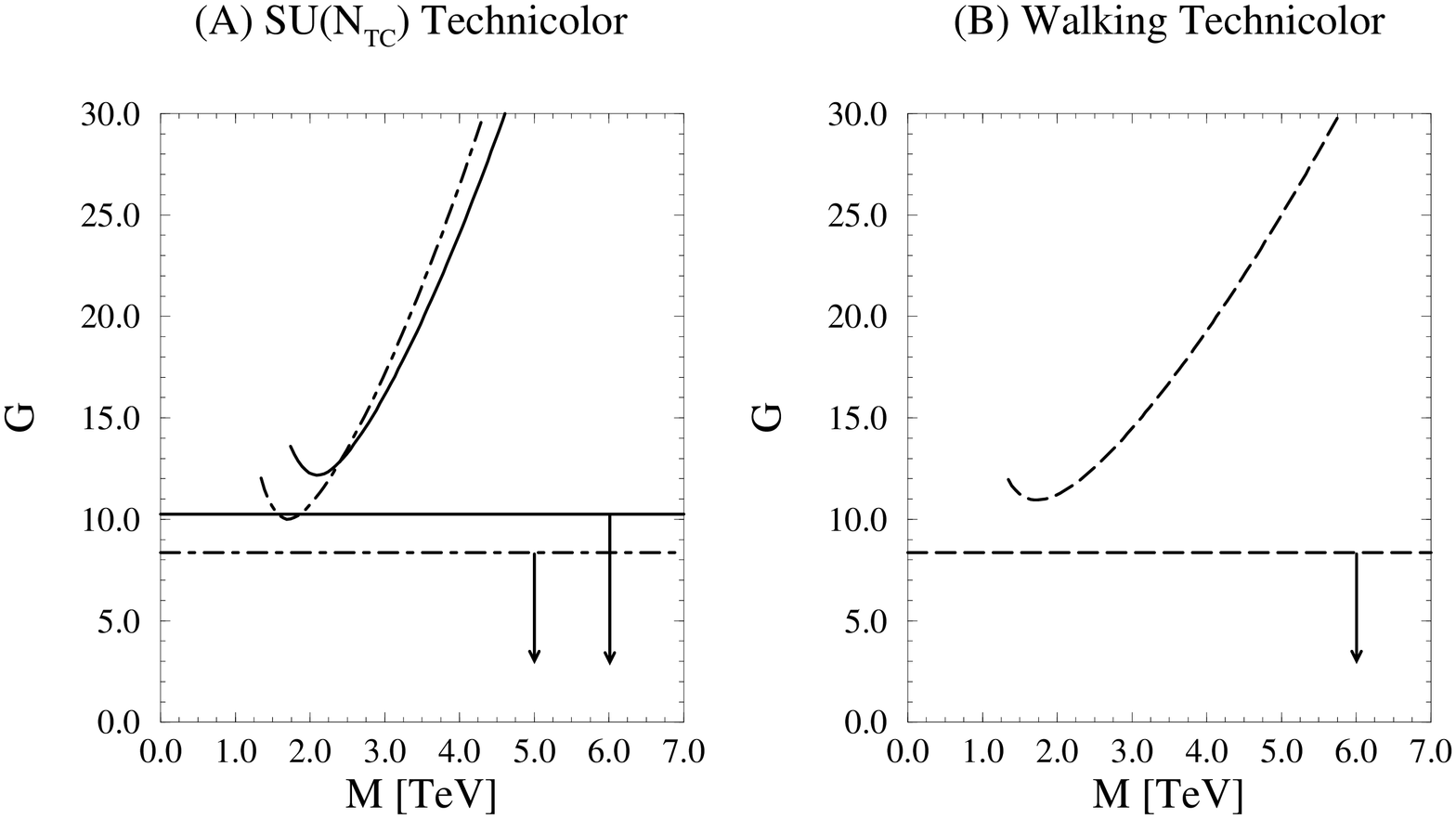,height=13cm,angle=270} }
\vspace{1cm}
{\Huge Fig. 3}
\end{center}
\end{figure}
\clearpage
\begin{figure}
\begin{center}
\centerline{ \epsfig{figure=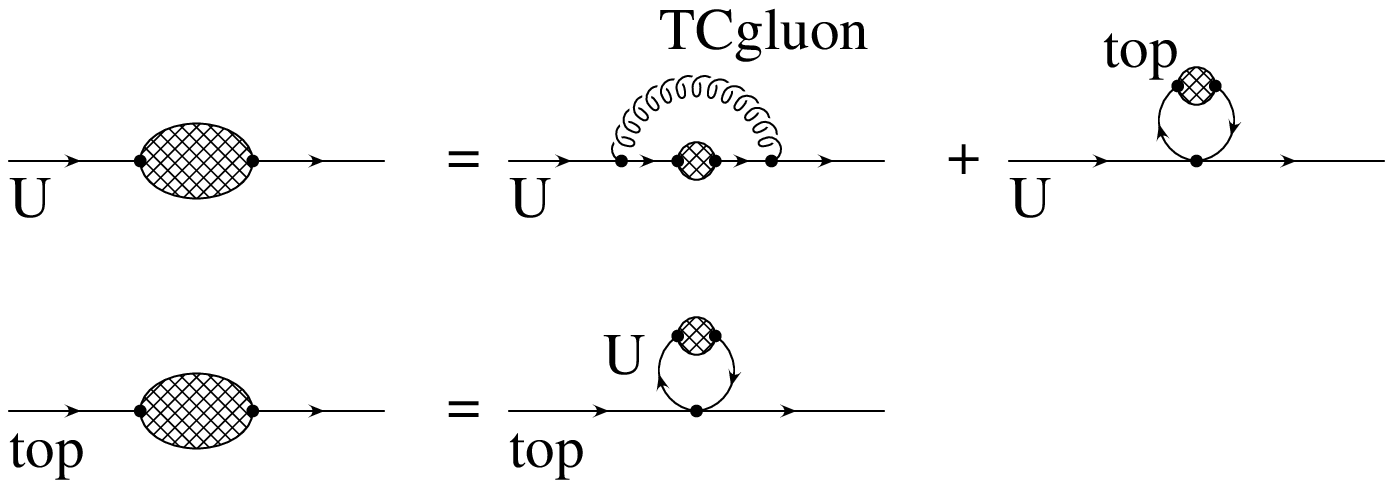,height=5cm} }
\vspace{1cm}
{\Huge Fig. 4}
\end{center}
\end{figure}
\clearpage
\begin{figure}
\begin{center}
\centerline{ \epsfig{figure=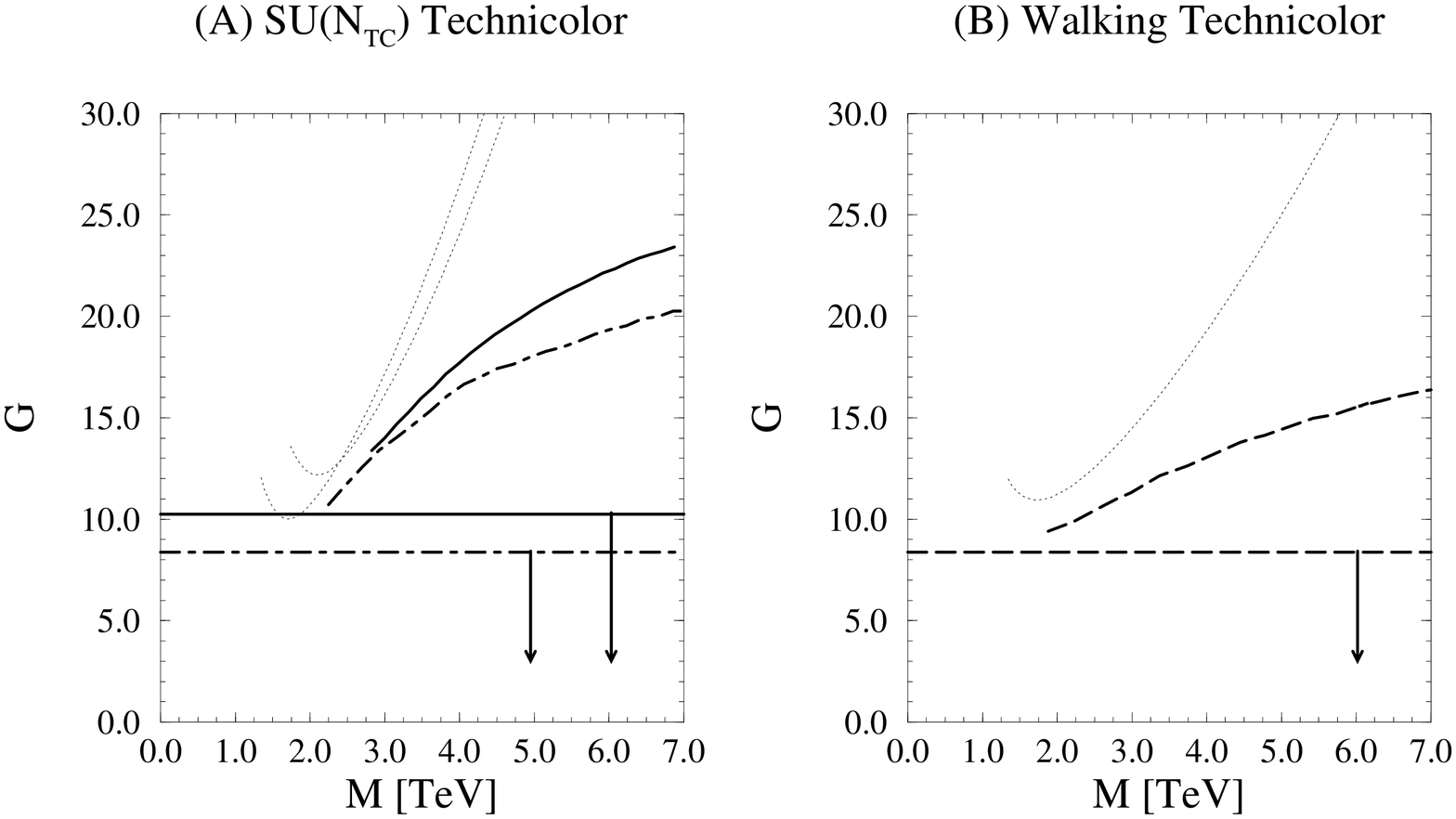,height=13cm,angle=270} }
\vspace{1cm}
{\Huge Fig. 5}
\end{center}
\end{figure}
\clearpage
\begin{figure}
\begin{center}
\centerline{ \epsfig{figure=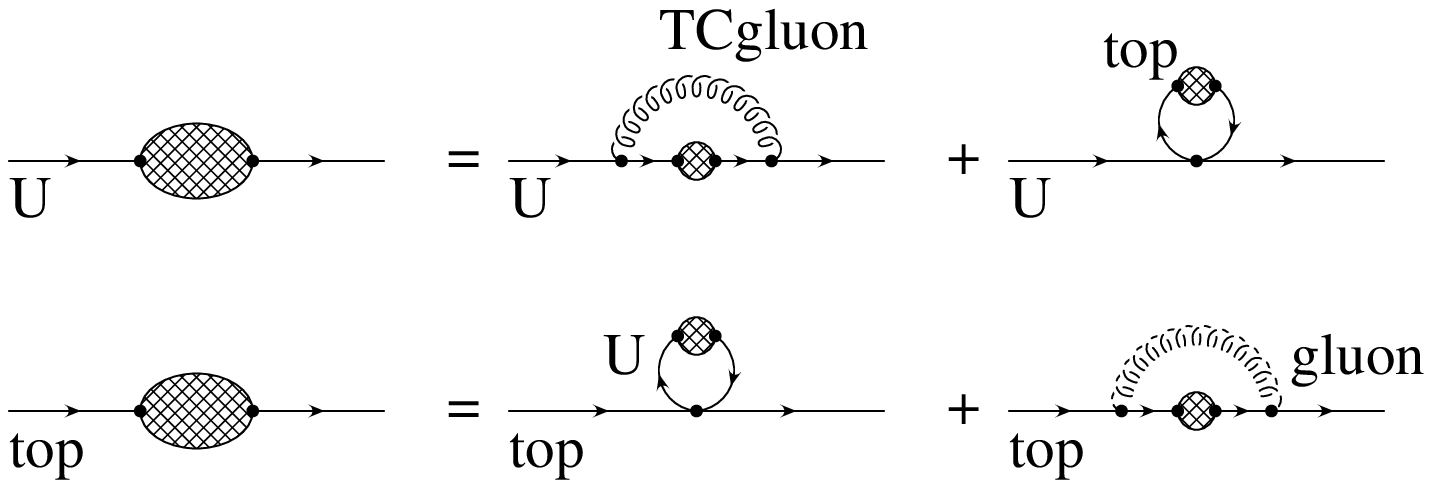,height=5cm} }
\vspace{1cm}
{\Huge Fig. 6}
\end{center}
\end{figure}
\clearpage
\begin{figure}
\begin{center}
\centerline{ \epsfig{figure=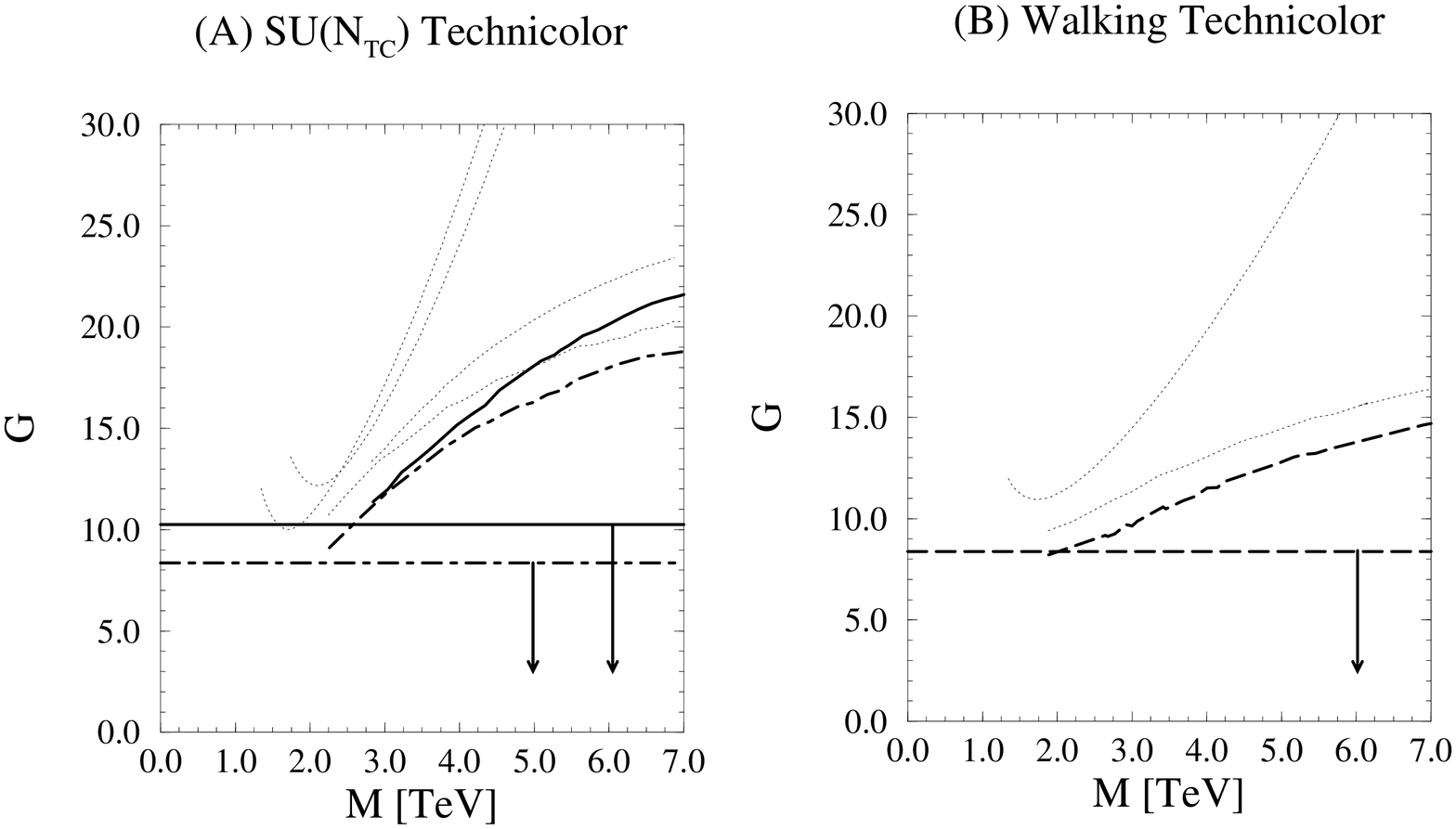,height=13cm, angle=270} }
\vspace{1cm}
{\Huge Fig. 7}
\end{center}
\end{figure}
\clearpage
\begin{figure}
\begin{center}
\centerline{
\epsfig{figure=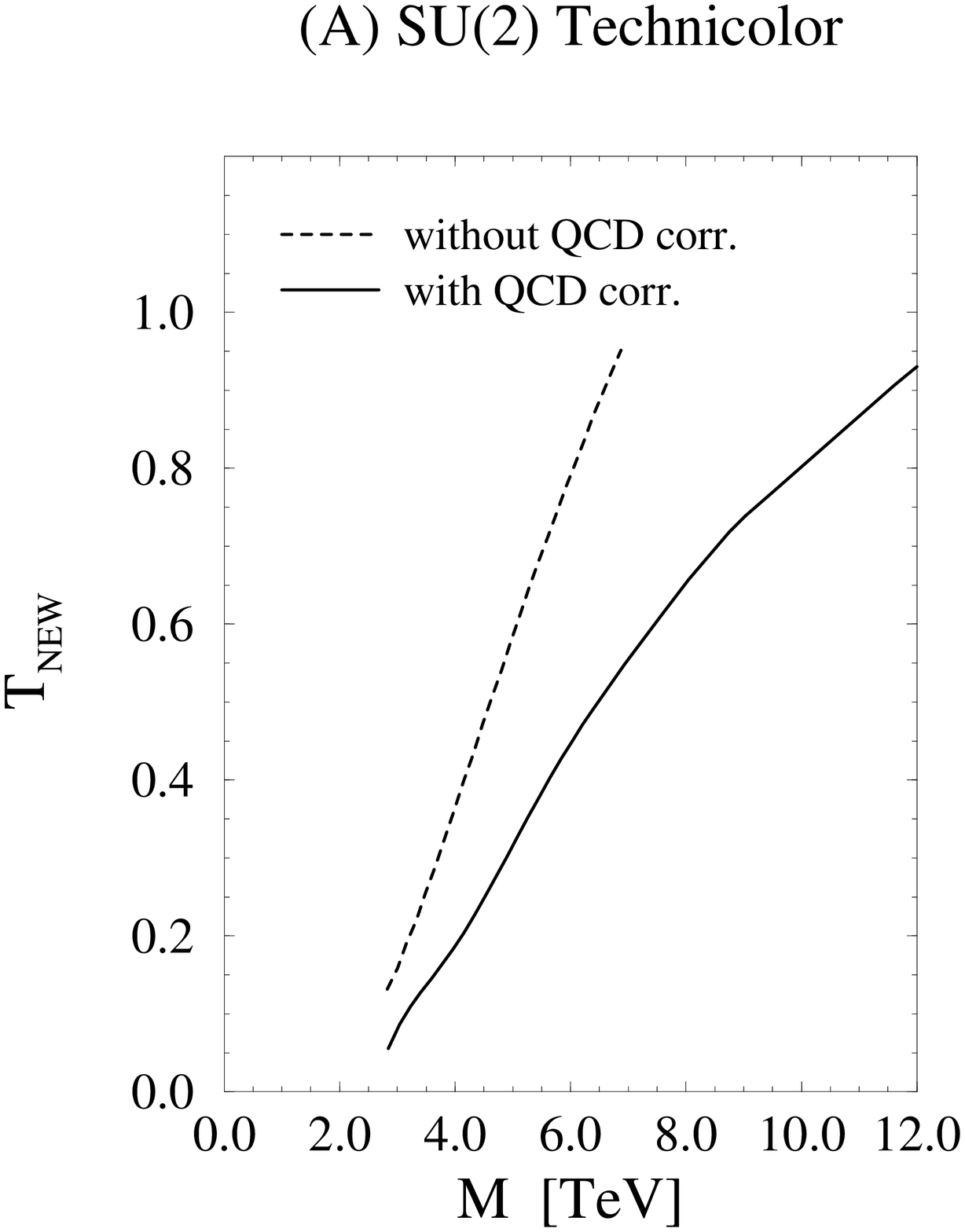,height=11cm, angle=0}%
\epsfig{figure=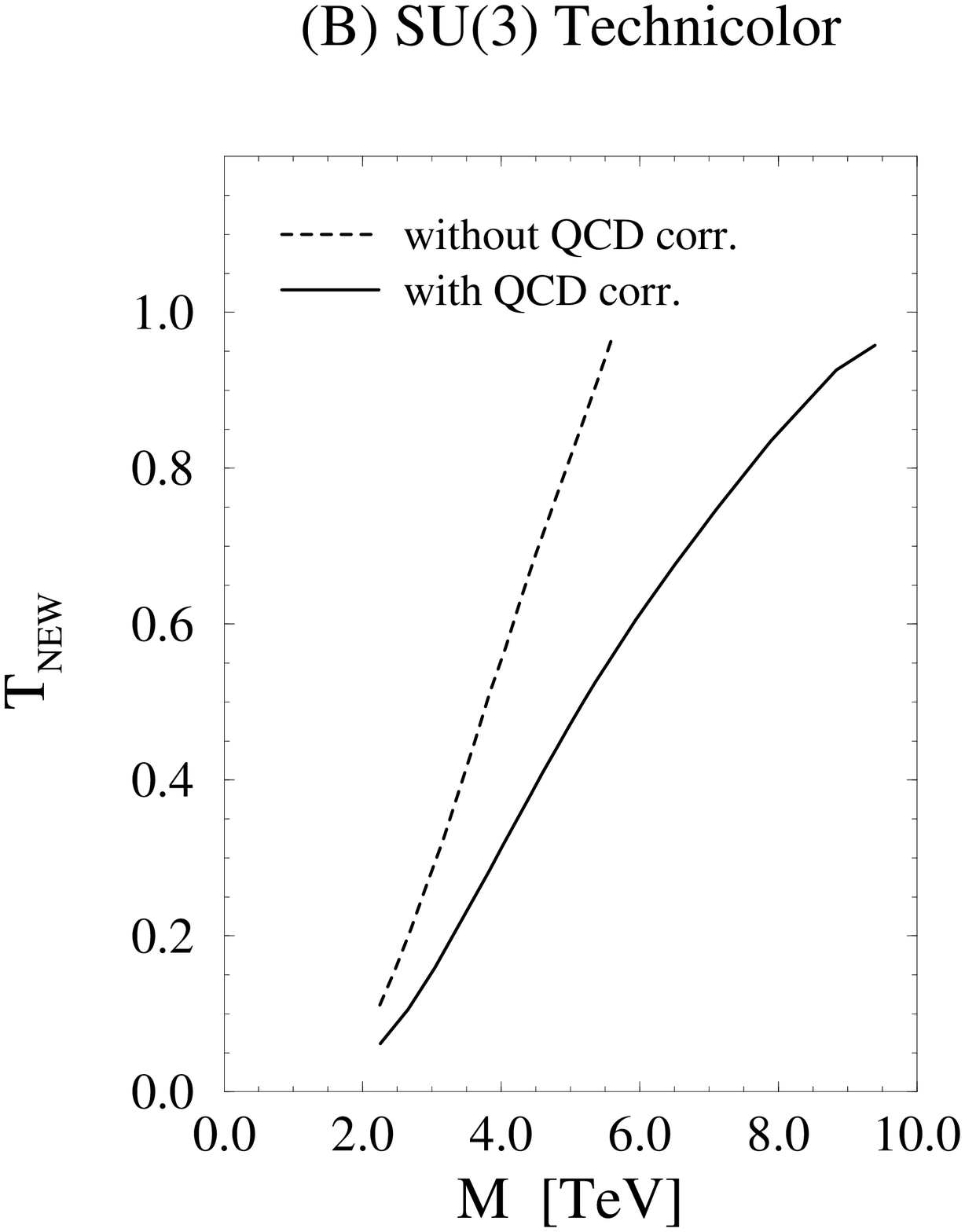,height=11cm, angle=0}
}
\vspace{1cm}
{\Huge Fig. 8-(A) and 8-(B)}
\end{center}
\end{figure}
\begin{figure}
\begin{center}
\centerline{\epsfig{figure=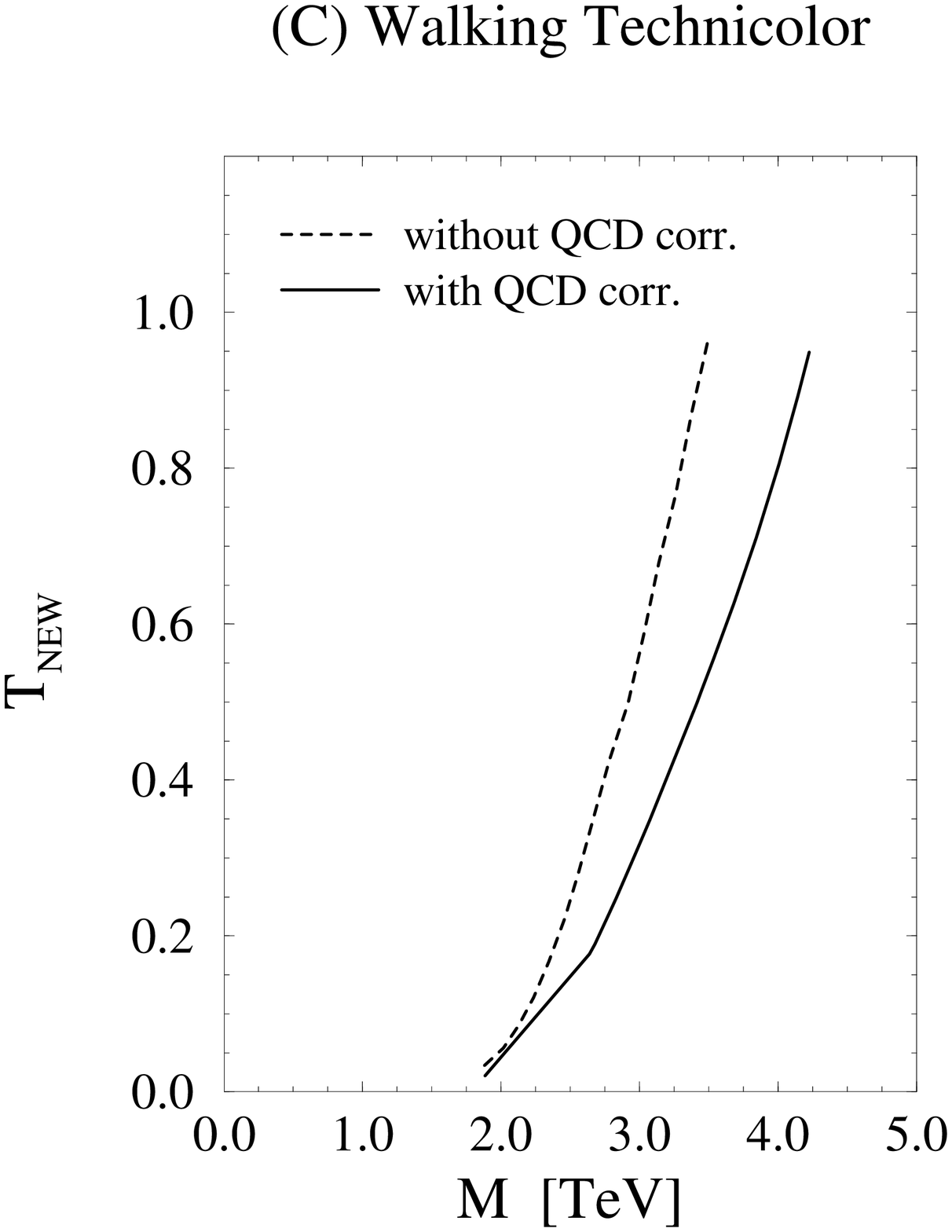,height=11cm, angle=0} }
\vspace{1cm}
{\Huge Fig. 8-(C)}
\end{center}
\end{figure}
\begin{figure}
\begin{center}
\centerline{\epsfig{figure=4fcoup.eps,height=5cm, angle=0}}
\vspace{1cm}
{\Huge Fig. 9}
\end{center}
\end{figure}
\begin{figure}
\begin{center}
\centerline{\epsfig{figure=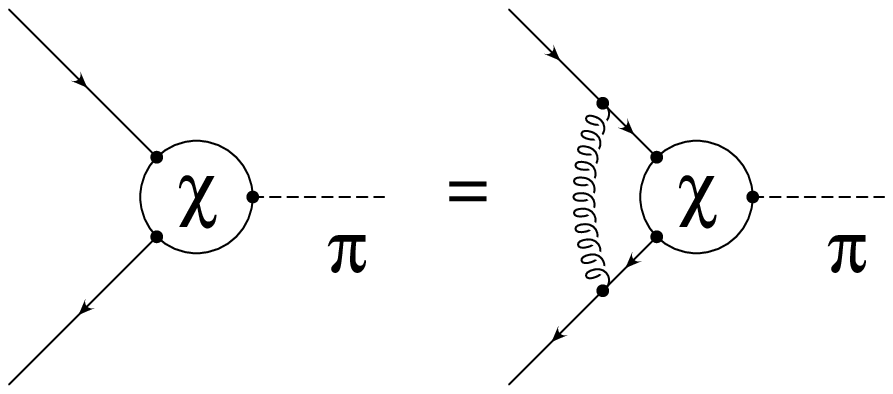,height=5cm, angle=0}}
\vspace{1cm}
{\Huge Fig. 10}
\end{center}
\end{figure}
%
%

\begin{thebibliography}{99}
\bibitem{cutoff}
More precisely in these equations we should set, respectively,
\bea
A_2 = 0, ~~~\Lambda = \infty, ~~~\mbox{and} 
~~~
\left\{
\begin{array}{ll}
A_1 = 1 & \mbox{for} ~~~x \le M^2 \nonumber\\
A_1 = 0 & \mbox{for} ~~~x \ge M^2,
\end{array}
\right.
\eea
and 
\bea
A_2 = 1, ~~~\Lambda = \infty, ~~~\mbox{and} 
~~~
\left\{
\begin{array}{ll}
A_1 = 1 & \mbox{for} ~~~x \le M^2 \nonumber\\
A_1 = 0 & \mbox{for} ~~~x \ge M^2.
\end{array}
\right.
\eea
Here, note that the four fermi operator (\ref{4fop}) resolves 
(and is crudely set equal to zero)
above the scale $M$. 
The mass functions of techni-fermions 
($\Sigma_U(x)$ and $\Sigma_D(x)$)
and also the running coupling constant $\lambda(x)$,
however, decrease in the energy region 
$x \gg \Lambda_{TC}^2$,
and we may replace the ultraviolet cut-off $\Lambda$ by $M$ 
approximately.
Thus we have only to know the mass fuctions below the scale $M$
and solve the coupled equations Eqs.(\ref{ap-csdeq}) 
and (\ref{ap-csdqcdeq}).
In fact, we see in Figs.\ref{fig-gmcsd} that
the obtained results from the coupled Schwinger-Dyson equations
Eq.(\ref{ap-csdeq}) are close to those from the 
Schwinger-Dyson equation Eq.(\ref{ap-sdeq}) as
the mass scale $M$ approaches $\Lambda_{TC}$.
Thus we consider this approximation does not change 
our results.
\end{thebibliography}
\end{document}